 \documentclass[final,5p,times,twocolumn]{elsarticle}

\usepackage{graphicx}
\usepackage{amssymb}
 \usepackage{lineno}

\journal{NIM-A}

\begin{document}

\begin{frontmatter}



\author[NIST]{Richard M. Lindstrom}
\address[NIST]{Analytical Chemistry Division National Institute of Standards and Technology, Gaithersburg, MD 20899, USA}
\author[PurduePhys]{Ephraim Fischbach\corref{cor1}}
\address[PurduePhys]{Department of Physics, Purdue University, West Lafayette, IN 47907 USA}
\cortext[cor1]{Corresponding author}
\ead{ephraim@purdue.edu}
\author[Witt]{John B. Buncher}
\address[Witt]{Department of Physics, Wittenberg University, Springfield, Ohio 45501, USA}
\author[PurdueNE]{Jere H. Jenkins}
\address[PurdueNE]{School of Nuclear Engineering, Purdue University, 400 Central Dr., West Lafayette, IN  47907 USA}
\author[UTK]{Andrew Yue}
\address[UTK]{Department of Physics, University of Tennessee, Knoxville, TN 37996, USA}

%%% \author[PurdueNE,PurduePhys]{Jere H. Jenkins\corref{cor1}}
%%% \address[PurdueNE]{School of Nuclear Engineering, Purdue University, 400 Central Dr., West Lafayette, IN  47907 USA}
%%% \address[PurduePhys]{Department of Physics, Purdue University, West Lafayette, IN 47907}
%%% \ead{jere@purdue.edu}
%%% \cortext[cor1]{Corresponding author}
%%% \author[OSURR]{Kevin R. Herminghuysen}
%%% \address[OSURR]{Ohio State University Research Reactor, The Ohio State University, Columbus, OH 43210 USA}
%%% \author[OSURR]{Thomas E. Blue}
%%% \author[PurduePhys]{Ephraim Fischbach}
%%% \author[Animal]{Daniel Javorsek II}
%%% \address[Animal]{411th Flight Test Squadron, 412th Test Wing, Edwards AFB, CA 93524, USA}
%%% \author[OSURR]{Andrew C. Kauffman}
%%% \author[Mayo]{Daniel W. Mundy}
%%% \address[Mayo]{Department of Radiation Oncology Physics, Mayo Clinic, Rochester, MN 55905 USA}
%%% \author[Stanford]{Peter A. Sturrock}
%%% \address[Stanford]{Center for Space Science and Astrophysics, Stanford University, Stanford, CA 94305 USA}
%%% \author[OSURR]{Joseph A. Talnagi}

\title{Absence of a self-induced decay effect in $^{198}$Au}


%\author{}
%
%\address{}

\begin{abstract}
We report the results of an improved experiment aimed at determining whether the half-life ($T_{1/2}$) of $^{198}$Au depends on the shape of the source. In this experiment, the half-lives of a gold sphere and a thin gold wire were measured after each had been irradiated in the NIST Center for Neutron Research. In comparison to an earlier version of this experiment, both the specific activities of the samples and their relative surface/volume ratios have been increased, leading to an improved test for the hypothesized self-induced decay (SID) effect. We find $T_{1/2(sphere)}/T_{1/2(wire)} = (0.9993\pm0.0002)$, which is compatible with no SID effect.
\end{abstract}


\end{frontmatter}


\section{Introduction}
\label{Intro}

In the first 35 years after Becquerel's discovery of radioactivity, more than eighty attempts were made to alter the rate of decay of a radioactive source by artificial means. Temperatures from  -255 to 1350$^{\circ}$C (18 to 1600 K), pressures up to 200 MPa (2000 atm), accelerations up to nearly 10$^7$ m$^2$/s (970,000 g), and magnetic fields as high as 8.3 T have all been found to be ineffectual \citep{hah76}. External gamma radiation also had no effect \citep{hev22}, and recent suggestions of a change at very low temperature in metals have not been confirmed \citep{rup08,har10}. The sole verified exception to constant decay rates is the influence of chemical state upon decay modes involving atomic electrons, such as electron capture and internal conversion, the physics of which is well understood \citep{eme72,dan79}. 

It has recently been suggested \citep{jen09a,jen09b,fis09,stu10a,jav10,stu10b,stu11a,stu11b} that short-duration and annual periodic anomalies, of order $\pm$0.3\%, that have been observed in several cases of decay of radioactive sources may be related to the Sun, specifically to solar neutrinos. Although neutrinos have been suggested to exhibit remarkable physics \citep{web88}, correlations of radioactive decay with the Earth-Sun distance have not been found in other data sets \citep{nor09} at the 0.3\% level, although they do appear to be present in these data at a lower level \cite{oke11}. Constraints on electron anti-neutrinos as a source of the periodic signals observed in various data sets have been derived by de Meijer et al., from a recent experiment using a research reactor as a source of neutrinos \cite{dem11}. Previous work in this laboratory \citep{lin10} compared the decay constants of spherical and foil sources of $^{198}$Au, with very different surface/volume ratios and hence very different internal neutrino density, and found no detectable difference in the half-life. In the present work, we have extended these measurements by doubling both the specific activity and the surface/volume ratio difference of the earlier work. 

\section{Experimental}
\label{Exp}

A 1.004 mg sphere, prepared by melting gold foil, was compared with a 33 $\mu$m diameter gold wire 6.0 cm long, weighing 1.02 mg. Each sample was irradiated for six hours (the maximum permissible time) in the RT1 rabbit facility of the NIST Center for Neutron Research, at a neutron flux of 1.05 x 10$^{14}$ n/cm$^2$s. Beginning a few hours after the irradiation ended, the source was counted repeatedly for 2 h live time through about 2.5 cm of lead absorber at 40 cm from a well-characterized Ge gamma detector (relative efficiency 37.3\%, resolution 1.70 keV), collecting $\sim$200 spectra during seven half-lives. A precision pulser \citep{the97,lin05} monitored the rate-related spectrometer losses. At the beginning of the measurement, each source had an activity of 62 mCi (2.3 GBq), emitting 2.3 x 10$^{12}$ neutrinos/s$\cdot$g. 

The 412-keV gamma ray peak in each spectrum was integrated with a fixed-boundary summation routine \citep{lin94}. Each datum was corrected for decay during the dead-time-extended counting interval, and for pulse pileup using the reference pulser. The data set was then fitted to an exponential function by a nonlinear reduced gradient method for $\chi^2$ minimization, using the Solver function \citep{fyl98} in Microsoft Excel to determine the half-life ($T_{1/2}$) and the initial activity ($A_o$). The weight of each data point was taken as the inverse Poisson variance of the net peak areas plus that of the missing pulser counts. Fits were performed over the entire data sets, as well as subsets comprising the first and last half-lives.

\section{Results and Discussion}
\label{Results}

The results are given in Table 1. The uncertainty $s$ is the increment in $T_{1/2}$ that increases the value of $\chi^2/df$ by unity, where $df$ is the number of data points minus 2. It is evident that not all sources of uncertainty are accounted for. Since approximately 4$\times$10$^6$ net counts were accumulated in the first spectrum, the very small Poisson uncertainty in this number is likely to be overshadowed by small, incompletely understood non-linearities in high-rate gamma spectrometry, and thus the value of $\chi^2$ is higher than in our previous measurements. As with other published half-life measurements, it is arguable that the formal uncertainties are underestimates, and following the practice of the Particle Data Group \citep{nak10} should be inflated by a factor of $ \sqrt{\chi^2/df} $. Perhaps because this experiment was carried out under more extreme conditions, the absolute values of the half-life are about 0.5\% lower than our previous work and the consensus of a recent evaluation \citep{che11}.

Since the activity of the two sources at the end of irradiation differed by 14\% and the initial decay times were different, the comparison for the first half-life used spectra with comparable counting rates. A second model was fitted to the data, employing coupled radioactive decay and extending and non-extending dead times \citep{fle79,lin95}. This function is less facile to implement and gave slightly inferior $\chi^2$ values, but the conclusions are the same as with the simpler model.

\begin{table}

\label{tab:Table1}
\caption{Half-life of $^{198}$Au, comparing the full data set to a subset restricted to the data from the first half-life for each of the two sources (sphere, wire). $n$ is the number of data points included in each set.}
\footnotesize
%\begin{tabular}{|p{3cm}|p{2.5cm}|p{2.5cm}|p{2.5cm}|p{2.5cm}|p{2.5cm}|}
\begin{tabular}{llcccc}
\hline
 &  & $T_{1/2}$, h & $\pm{}s$ & $\chi^2/df$ & $n$ \\
\hline
All data, sphere  &  & 64.255 & 0.007 & 4.3 & 203 \\
All data, wire   &  & 64.298 & 0.007 & 2.8 & 201 \\
    & Sphere/wire & 0.9993 & 0.0002 &  &  \\

First $T_{1/2}$ only, sphere  &  & 64.290 & 0.056 & 6.6 & 26 \\
First $T_{1/2}$ only, wire   &  & 64.147 & 0.053 & 4.6 & 27 \\   
    & Sphere/wire & 1.0022 & 0.0012 &  &  \\    
\end{tabular} 
\end{table}

The mean solar neutrino flux at earth is 6.5$\times$10$^{10}$/cm$^2$s \citep{bah05}, varying according the 1/r$^2$ law by 6.9\% from perihelion to aphelion. The number of solar neutrinos passing through the surface (twice the projected area) of the gold sphere is then 1.1$\times$10$^8$/s, and through the wire 2.6$\times$10$^9$/s. By comparison, the number of internally generated neutrinos from $^{198}$Au passing through the surface of the sphere is 3.4$\times$10$^{11}$/s and through the surface of the wire 3.7$\times$10$^{10}$/s at the beginning of the measurements. The 20-MW NIST research reactor, located 49 m from the detector, contributed 1.3$\times$10$^{10}$ antineutrinos/cm$^2$s continuously during the experiment \citep{ber02}.

In the formalism of Ref. \citep{lin10}, the postulated effect of an internal flux of neutrinos is to modify the usual decay formula such that 

\begin{eqnarray}
\frac{dN\left(t\right)}{dt}=-\lambda_oN\left(t\right)\left[1+\xi\frac{N\left(t\right)}{N_o}\right],
\label{Eq:eq1}
\end{eqnarray}

\noindent{}where $\lambda_o=\rm{ln}2/T_{1/2}$ is the conventional $^{198}$Au decay constant. The constant $\xi$ is a phenomenological parameter that depends on the shape of the sample \citep{bun10}

\begin{eqnarray}
\xi=\frac{\eta{}N_o}{4\pi{}c}\left\langle\frac{1}{d^2}\right\rangle_V,
\label{Eq:eq2}
\end{eqnarray}

\noindent{}where  $\eta$ is a fundamental parameter characteristic of $^{198}$Au, and $\langle\dots\rangle_V$ denotes the average of 1/$d^2$  between two randomly chosen points separated by distance $d$ from a sample (sphere, wire, foil) of volume $V$. For the spherical sample of radius $R$,

\begin{eqnarray}
\left\langle\frac{1}{d^2}\right\rangle_{sphere} = \frac{9}{4R^2}=4.37\times10^7\rm{m}^{-2}.
\label{Eq:eq3}
\end{eqnarray}

\noindent{}The results for the wire and the previously measured foil \citep{lin10} were obtained numerically, and are given by 

\begin{eqnarray}
\left\langle\frac{1}{d^2}\right\rangle_{wire} = 0.54\left(2\right)\times10^7\rm{m}^{-2}
\label{Eq:eq4}
\end{eqnarray}

\begin{eqnarray}
\left\langle\frac{1}{d^2}\right\rangle_{foil} = 1.41\left(14\right)\times10^7\rm{m}^{-2}.
\label{Eq:eq5}
\end{eqnarray}

It follows that the use of a wire in the current experiment represents an increase in the internal neutrino flux by a factor of $\sim$2.6 compared to the foil used in Ref. \citep{lin10}. Interestingly, this geometric enhancement is larger than would be expected from a simple estimate based on the increase by a factor of $\sim$1.5 in the surface/volume ratio compared to the foil. Combining the results in Eqs. (\ref{Eq:eq3}-\ref{Eq:eq5}), we estimate that in the current experiment the internal antineutrino flux in the sphere is larger than that in the wire by a factor of 8.1. Since the activities in the current samples are $\sim$2.1 times greater than in the original experiment of Ref. \citep{lin10}, the net increase in the sensitivity of the current experiment to a putative SID effect is given by 

\begin{eqnarray}
2.1\frac{\left\langle\frac{1}{d^2}\right\rangle_{foil}}{\left\langle\frac{1}{d^2}\right\rangle_{wire}} \cong 5.5.
\label{Eq:eq6}
\end{eqnarray}

If the decay rate of the sources described by \citet{jen09b} had been due to a 6.9\% change in solar neutrino flux, then an order of magnitude difference in a much larger internal neutrino field in the present measurements should have been readily detectable in the decay rate. We can quantify the constraint that our results imply for the putative SID effect by using Equation 8 of Ref. \citep{lin10}.

\begin{eqnarray}
\frac{\left(\dot{N}_{sphere}/\dot{N}_{wire}\right)_{t=0}}{\left(\dot{N}_{sphere}/\dot{N}_{wire}\right)_{t\to\infty}}\cong 1+2\Delta\xi,
\label{Eq:eq7}
\end{eqnarray}

\noindent{}where $\dot{N}\equiv dN\left(t\right)/dt$, and $\Delta\xi=\xi\left(sphere\right)-\xi\left(wire\right)$. To evaluate the $\left(t=0\right)/\left(t\to{}\infty\right)$ ratio in Eq. (\ref{Eq:eq7}), we averaged the 412 keV data over one half-life at the beginning (end) of the experiment. We find $\Delta\xi=-0.010\pm0.006$, where the uncertainty is the standard uncertainty $u$. 

In conclusion, our results imply that the periodic signals reported in Refs. \citep{jen09a,jen09b,fis09,stu10a,jav10,stu10b,stu11a,stu11b} were probably not caused by the $\bar{\nu}_e$ component of solar neutrino flux. This conclusion thus supports, and is supported by, the previously cited reactor experiment of \citet{dem11}. However, this does not exclude the possibility that other components of solar neutrino flux $\left(\nu_e, \nu_\mu, \nu_\tau\right)$, or some other unknown particles, could be responsible for the effects reported in Refs. \citep{jen09a,jen09b,fis09,stu10a,jav10,stu10b,stu11a,stu11b}. Additionally, it may be that $^{198}$Au is simply less sensitive than other nuclides to whatever influences are responsible for the time-varying effects observed to date. As we have noted elsewhere, \citep{fis09} the sensitivity of radioactive nuclides to an external influence is likely to depend on decay energy and other kinematic effects, as well as on details of nuclear structure, and hence could vary from nuclide to nuclide.

\section*{Acknowledgements}

We thank R.F. Fleming, G.L. Greene, and D.E. Krause for useful discussions during the course of this work. Contributions of the National Institute of Standards and Technology are not subject to U.S. copyright. The work of EF is supported in part by USDOE contract No. DE-AC02-76ER071428.



%% References


%\bibliographystyle{model1a-num-names}
%\bibliography{ms}

\begin{thebibliography}{30}
\expandafter\ifx\csname natexlab\endcsname\relax\def\natexlab#1{#1}\fi
\providecommand{\bibinfo}[2]{#2}
\ifx\xfnm\relax \def\xfnm[#1]{\unskip,\space#1}\fi
%Type = Article
\bibitem[{Hahn et~al.(1976)Hahn, Born, and Kim}]{hah76}
\bibinfo{author}{H.-P. Hahn}, \bibinfo{author}{H.-J. Born},
  \bibinfo{author}{J.~Kim}, \bibinfo{journal}{Radiochim. Acta}
  \bibinfo{volume}{23} (\bibinfo{year}{1976}) \bibinfo{pages}{23}.
%Type = Article
\bibitem[{Hevesy(1922)}]{hev22}
\bibinfo{author}{G.~Hevesy}, \bibinfo{journal}{Nature} \bibinfo{volume}{110}
  (\bibinfo{year}{1922}).
%Type = Article
\bibitem[{Ruprecht et~al.(2008)Ruprecht, Vockenhuber, Buchmann, Woods, Ruiz,
  Lapi, and Bemmerer}]{rup08}
\bibinfo{author}{G.~Ruprecht}, \bibinfo{author}{C.~Vockenhuber},
  \bibinfo{author}{L.~Buchmann}, \bibinfo{author}{R.~Woods},
  \bibinfo{author}{C.~Ruiz}, \bibinfo{author}{S.~Lapi},
  \bibinfo{author}{D.~Bemmerer}, \bibinfo{journal}{Phys. Rev. C}
  \bibinfo{volume}{77} (\bibinfo{year}{2008}) \bibinfo{pages}{065502}.
%Type = Article
\bibitem[{Hardy et~al.(2010)Hardy, Goodwin, Golovko, and Iacob}]{har10}
\bibinfo{author}{J.~C. Hardy}, \bibinfo{author}{J.~R. Goodwin},
  \bibinfo{author}{V.~V. Golovko}, \bibinfo{author}{V.~E. Iacob},
  \bibinfo{journal}{Appl. Radiat. Isot.} \bibinfo{volume}{68}
  (\bibinfo{year}{2010}) \bibinfo{pages}{1550}.
%Type = Article
\bibitem[{Emery(1972)}]{eme72}
\bibinfo{author}{G.~T. Emery}, \bibinfo{journal}{Ann. Rev. Nucl. Sci.}
  \bibinfo{volume}{22} (\bibinfo{year}{1972}) \bibinfo{pages}{165--202}.
%Type = Article
\bibitem[{Daniel(1979)}]{dan79}
\bibinfo{author}{H.~Daniel}, \bibinfo{journal}{At. Energy Rev.}
  \bibinfo{volume}{17} (\bibinfo{year}{1979}) \bibinfo{pages}{287}.
%Type = Article
\bibitem[{Jenkins and Fischbach(2009)}]{jen09a}
\bibinfo{author}{J.~Jenkins}, \bibinfo{author}{E.~Fischbach},
  \bibinfo{journal}{Astropart. Phys.} \bibinfo{volume}{31}
  (\bibinfo{year}{2009}) \bibinfo{pages}{407}.
%Type = Article
\bibitem[{Jenkins et~al.(2009)Jenkins, Fischbach, Buncher, Gruenwald, Krause,
  and Mattes}]{jen09b}
\bibinfo{author}{J.~Jenkins}, \bibinfo{author}{E.~Fischbach},
  \bibinfo{author}{J.~Buncher}, \bibinfo{author}{J.~Gruenwald},
  \bibinfo{author}{D.~Krause}, \bibinfo{author}{J.~Mattes},
  \bibinfo{journal}{Astropart. Phys.} \bibinfo{volume}{32}
  (\bibinfo{year}{2009}) \bibinfo{pages}{42}.
%Type = Article
\bibitem[{Fischbach et~al.(2009)Fischbach, Buncher, Gruenwald, Jenkins, Krause,
  Mattes, and Newport}]{fis09}
\bibinfo{author}{E.~Fischbach}, \bibinfo{author}{J.~Buncher},
  \bibinfo{author}{J.~Gruenwald}, \bibinfo{author}{J.~Jenkins},
  \bibinfo{author}{D.~Krause}, \bibinfo{author}{J.~Mattes},
  \bibinfo{author}{J.~Newport}, \bibinfo{journal}{Space Sci. Rev.}
  \bibinfo{volume}{145} (\bibinfo{year}{2009}) \bibinfo{pages}{285}.
%Type = Article
\bibitem[{Sturrock et~al.(2010)Sturrock, Buncher, Fischbach, Gruenwald,
  Javorsek~II, Jenkins, Lee, Mattes, and Newport}]{stu10a}
\bibinfo{author}{P.~A. Sturrock}, \bibinfo{author}{J.~B. Buncher},
  \bibinfo{author}{E.~Fischbach}, \bibinfo{author}{J.~T. Gruenwald},
  \bibinfo{author}{D.~Javorsek~II}, \bibinfo{author}{J.~H. Jenkins},
  \bibinfo{author}{R.~H. Lee}, \bibinfo{author}{J.~J. Mattes},
  \bibinfo{author}{J.~R. Newport}, \bibinfo{journal}{Astropart. Phys.}
  \bibinfo{volume}{34} (\bibinfo{year}{2010}) \bibinfo{pages}{121}.
%Type = Article
\bibitem[{Javorsek~II et~al.(2010)Javorsek~II, Sturrock, Lasenby, Lasenby,
  Buncher, Fischbach, Gruenwald, Hoft, Horan, Jenkins, Kerford, Lee, Longman,
  Mattes, Morreale, Morris, Mudry, Newport, O'Keefe, Petrelli, Silver, Stewart,
  and Terry}]{jav10}
\bibinfo{author}{D.~Javorsek~II}, \bibinfo{author}{P.~A. Sturrock},
  \bibinfo{author}{R.~N. Lasenby}, \bibinfo{author}{A.~N. Lasenby},
  \bibinfo{author}{J.~B. Buncher}, \bibinfo{author}{E.~Fischbach},
  \bibinfo{author}{J.~T. Gruenwald}, \bibinfo{author}{A.~W. Hoft},
  \bibinfo{author}{T.~J. Horan}, \bibinfo{author}{J.~H. Jenkins},
  \bibinfo{author}{J.~L. Kerford}, \bibinfo{author}{R.~H. Lee},
  \bibinfo{author}{A.~Longman}, \bibinfo{author}{J.~J. Mattes},
  \bibinfo{author}{B.~L. Morreale}, \bibinfo{author}{D.~B. Morris},
  \bibinfo{author}{R.~N. Mudry}, \bibinfo{author}{J.~R. Newport},
  \bibinfo{author}{D.~O'Keefe}, \bibinfo{author}{M.~A. Petrelli},
  \bibinfo{author}{M.~A. Silver}, \bibinfo{author}{C.~A. Stewart},
  \bibinfo{author}{B.~Terry}, \bibinfo{journal}{Astropart. Phys.}
  \bibinfo{volume}{34} (\bibinfo{year}{2010}) \bibinfo{pages}{173}.
%Type = Article
\bibitem[{Sturrock et~al.(2010)Sturrock, Buncher, Fischbach, Gruenwald,
  Javorsek, Jenkins, Lee, Mattes, and Newport}]{stu10b}
\bibinfo{author}{P.~Sturrock}, \bibinfo{author}{J.~Buncher},
  \bibinfo{author}{E.~Fischbach}, \bibinfo{author}{J.~Gruenwald},
  \bibinfo{author}{D.~Javorsek}, \bibinfo{author}{J.~Jenkins},
  \bibinfo{author}{R.~Lee}, \bibinfo{author}{J.~Mattes},
  \bibinfo{author}{J.~Newport}, \bibinfo{journal}{Sol. Phys.}
  \bibinfo{volume}{267} (\bibinfo{year}{2010}) \bibinfo{pages}{251}.
%Type = Article
\bibitem[{Sturrock et~al.(2011)Sturrock, Fischbach, and Jenkins}]{stu11a}
\bibinfo{author}{P.~Sturrock}, \bibinfo{author}{E.~Fischbach},
  \bibinfo{author}{J.~Jenkins}, \bibinfo{journal}{Sol. Phys.}
  (\bibinfo{year}{2011}) \bibinfo{pages}{In Press}. [doi:10.1007/s11207-011-9807-5]
%Type = Article
\bibitem[{Sturrock et~al.(2011)Sturrock, Buncher, Fischbach, Javorsek~II,
  Jenkins, and Mattes}]{stu11b}
\bibinfo{author}{P.~Sturrock}, \bibinfo{author}{J.~Buncher},
  \bibinfo{author}{E.~Fischbach}, \bibinfo{author}{D.~Javorsek~II},
  \bibinfo{author}{J.~H. Jenkins}, \bibinfo{author}{J.~J. Mattes},
  \bibinfo{journal}{ApJ} (\bibinfo{year}{2011}) \bibinfo{pages}{In Press}. [doi:10.1088/0004-637X/735/1/1]
%Type = Article
\bibitem[{Weber(1988)}]{web88}
\bibinfo{author}{J.~Weber}, \bibinfo{journal}{Phys. Rev. D}
  \bibinfo{volume}{38} (\bibinfo{year}{1988}) \bibinfo{pages}{32}.
%Type = Article
\bibitem[{Norman et~al.(2009)Norman, Browne, Shugart, Joshi, and
  Firestone}]{nor09}
\bibinfo{author}{E.~Norman}, \bibinfo{author}{E.~Browne},
  \bibinfo{author}{H.~Shugart}, \bibinfo{author}{T.~Joshi},
  \bibinfo{author}{R.~Firestone}, \bibinfo{journal}{Astropart. Phys.}
  \bibinfo{volume}{31} (\bibinfo{year}{2009}) \bibinfo{pages}{135}.
%Type = Article
\bibitem[{O'Keefe et~al.(2011)O'Keefe, Morreale, Fischbach, Javorsek~II,
  Jenkins, Lee, Morris, and Sturrock}]{oke11}
\bibinfo{author}{D.~O'Keefe}, \bibinfo{author}{B.~Morreale},
  \bibinfo{author}{E.~Fischbach}, \bibinfo{author}{D.~Javorsek~II},
  \bibinfo{author}{J.~Jenkins}, \bibinfo{author}{R.~H. Lee},
  \bibinfo{author}{D.~Morris}, \bibinfo{author}{P.~Sturrock}
  (\bibinfo{year}{2011 in preparation.}).
%Type = Article
\bibitem[{de~Meijer et~al.(2011)de~Meijer, Blaauw, and Smit}]{dem11}
\bibinfo{author}{R.~J. de~Meijer}, \bibinfo{author}{M.~Blaauw},
  \bibinfo{author}{F.~D. Smit}, \bibinfo{journal}{Appl. Radiat. Isot.}
  \bibinfo{volume}{69} (\bibinfo{year}{2011}) \bibinfo{pages}{320}.
%Type = Article
\bibitem[{Lindstrom et~al.(2010)Lindstrom, Fischbach, Buncher, Greene, Jenkins,
  Krause, Mattes, and Yue}]{lin10}
\bibinfo{author}{R.~M. Lindstrom}, \bibinfo{author}{E.~Fischbach},
  \bibinfo{author}{J.~B. Buncher}, \bibinfo{author}{G.~L. Greene},
  \bibinfo{author}{J.~H. Jenkins}, \bibinfo{author}{D.~E. Krause},
  \bibinfo{author}{J.~J. Mattes}, \bibinfo{author}{A.~Yue},
  \bibinfo{journal}{Nucl. Instrum. Methods Phys. Res., Sect. A}
  \bibinfo{volume}{622} (\bibinfo{year}{2010}) \bibinfo{pages}{93}.
%Type = Article
\bibitem[{Then et~al.(1997)Then, Geurink, Bode, and Lindstrom}]{the97}
\bibinfo{author}{S.~Then}, \bibinfo{author}{F.~Geurink},
  \bibinfo{author}{P.~Bode}, \bibinfo{author}{R.~Lindstrom},
  \bibinfo{journal}{J. Radioanal. Nucl. Chem.} \bibinfo{volume}{215}
  (\bibinfo{year}{1997}) \bibinfo{pages}{249}.
%Type = Article
\bibitem[{Lindstrom et~al.(2005)Lindstrom, Blaauw, and Unterweger}]{lin05}
\bibinfo{author}{R.~M. Lindstrom}, \bibinfo{author}{M.~Blaauw},
  \bibinfo{author}{M.~P. Unterweger}, \bibinfo{journal}{J. Radioanal. Nucl.
  Chem.} \bibinfo{volume}{263} (\bibinfo{year}{2005}) \bibinfo{pages}{311}.
%Type = Article
\bibitem[{Lindstrom(1994)}]{lin94}
\bibinfo{author}{R.~Lindstrom}, \bibinfo{journal}{Biol. Trace Elem. Res.}
  \bibinfo{volume}{43-45} (\bibinfo{year}{1994}) \bibinfo{pages}{597}.
%Type = Article
\bibitem[{Fylstra et~al.(1998)Fylstra, Lasdon, Watson, and Waren}]{fyl98}
\bibinfo{author}{D.~Fylstra}, \bibinfo{author}{L.~Lasdon},
  \bibinfo{author}{J.~Watson}, \bibinfo{author}{A.~Waren},
  \bibinfo{journal}{Interfaces} \bibinfo{volume}{28} (\bibinfo{year}{1998})
  \bibinfo{pages}{29}.
%Type = Article
\bibitem[{Nakamura and Particle~Data(2010)}]{nak10}
\bibinfo{author}{K.~Nakamura}, \bibinfo{author}{G.~Particle~Data},
  \bibinfo{journal}{J. Phys. G: Nucl. Part. Phys.} \bibinfo{volume}{37}
  (\bibinfo{year}{2010}) \bibinfo{pages}{075021}.
%Type = Article
\bibitem[{Chen et~al.(2011)Chen, Geraedts, Ouellet, and Singh}]{che11}
\bibinfo{author}{J.~Chen}, \bibinfo{author}{S.~Geraedts},
  \bibinfo{author}{C.~Ouellet}, \bibinfo{author}{B.~Singh},
  \bibinfo{journal}{Appl. Radiat. Isot.} \bibinfo{volume}{69}
  (\bibinfo{year}{2011}) \bibinfo{pages}{1064}.
%Type = Inproceedings
\bibitem[{Fleming(1979)}]{fle79}
\bibinfo{author}{R.~F. Fleming}, in: \bibinfo{booktitle}{Trans. Am. Nucl. Soc. 1979 Winter Meeting, 11-15 Nov. 1979},
  Volume~\bibinfo{volume}{33} of \textit{\bibinfo{series}{Trans. Am. Nucl. Soc.
  }}, \bibinfo{address}{USA}, pp. \bibinfo{pages}{234--5}.
%Type = Article
\bibitem[{Lindstrom and Fleming(1995)}]{lin95}
\bibinfo{author}{R.~M. Lindstrom}, \bibinfo{author}{R.~F. Fleming},
  \bibinfo{journal}{Radioactivity \& Radiochemistry} \bibinfo{volume}{6}
  (\bibinfo{year}{1995}) \bibinfo{pages}{20}.
%Type = Article
\bibitem[{Bahcall et~al.(2005)Bahcall, Serenelli, and Basu}]{bah05}
\bibinfo{author}{J.~N. Bahcall}, \bibinfo{author}{A.~M. Serenelli},
  \bibinfo{author}{S.~Basu}, \bibinfo{journal}{ApJ Lett.} \bibinfo{volume}{621}
  (\bibinfo{year}{2005}) \bibinfo{pages}{85}.
%Type = Article
\bibitem[{Bernstein et~al.(2002)Bernstein, Wang, Gratta, and West}]{ber02}
\bibinfo{author}{A.~Bernstein}, \bibinfo{author}{Y.~Wang},
  \bibinfo{author}{G.~Gratta}, \bibinfo{author}{T.~West}, \bibinfo{journal}{J.
  Appl. Phys.} \bibinfo{volume}{91} (\bibinfo{year}{2002})
  \bibinfo{pages}{4672}.
%Type = Phdthesis
\bibitem[{Buncher(2010)}]{bun10}
\bibinfo{author}{J.~B. Buncher}, \bibinfo{title}{Phenomenology of Time-varying
  Nuclear Decay Parameters}, \bibinfo{type}{Ph.D.}, Purdue University,
  \bibinfo{address}{West Lafayette, Indiana, USA}, \bibinfo{year}{2010}.

\end{thebibliography}



\end{document}